\documentclass[aps,prl,twocolumn,superscriptaddress,reprint]{revtex4-1}
\usepackage{graphicx}
\usepackage{setspace}
\usepackage{amsmath}
\usepackage{amssymb}
\usepackage{color}
\usepackage{array}
\usepackage{subfigure}
\usepackage{hyperref}
\usepackage{float}
\usepackage{lipsum}
\usepackage{ulem}

\usepackage[all]{xy}
\newcommand{\RN}[1]{%
  \textup{\uppercase\expandafter{\romannumeral#1}}%
}

\begin{document}
\title{Quantum simulation of the Weyl equation with a trapped ion }
\author{De-Sheng Li}
\email[email: ]{lideshengjy@126.com}
\affiliation{Interdisciplinary Center for Quantum Information, National University of Defense Technology, Changsha 410073, P.R.China}

\author{Chun-Wang Wu}
\email[email: ]{cwwu@nudt.edu.cn}
\affiliation{Interdisciplinary Center for Quantum Information, National University of Defense Technology, Changsha 410073, P.R.China}

\author{Lin-Ze He}
\email[email: ]{helinze12@nudt.edu.cn}
\affiliation{Interdisciplinary Center for Quantum Information, National University of Defense Technology, Changsha 410073, P.R.China}

\author{Wei Wu}
\email[email: ]{weiwu@nudt.edu.cn}
\affiliation{Interdisciplinary Center for Quantum Information, National University of Defense Technology, Changsha 410073, P.R.China}

\author{Ping-Xing Chen}
\email[email: ]{pxchen@nudt.edu.cn}
\affiliation{Interdisciplinary Center for Quantum Information, National University of Defense Technology, Changsha 410073, P.R.China}

\begin{abstract}
The Weyl equation describes chiral massless relativistic particles, called Weyl fermions, which have important relations to neutrinos. A direct observation of the dynamics of Weyl fermions in an experiment is difficult to achieve. This study investigates a method of simulating the Weyl equation in $1+2$-dimension by a single trapped ion. The predictions about a two-dimensional Zitterbewegung and an especially interesting phenomenon of Weyl fermions can be tested by the future trapped ion experiment, which might enhance our understanding of neutrinos.
\end{abstract}
\maketitle


\section{Introduction}
\label{intro}
The Weyl equation can be derived from the Dirac equation as rest mass equals zero. Neutrinos have a nonzero tiny mass; hence, the Weyl equation can be used to approximately describe neutrinos. In the Standard Model of particle physics (SM), neutrinos are assumed to be massless, and only left-handed neutrinos exist \cite{patrignani2016review}. 
The following remains an open question: why do left-handed neutrinos exist, but not right-handed neutrinos? Neutrinos do not have an electric charge, and are difficult to detect. Moreover, its properties are still not exactly clear. 
An analog quantum simulation \cite{Cirac2012goals,Blatt2012Quantum,Georgescu2013Quantum,Georgescu2014Quantum,Arrazola2016digital-analog} of the Weyl equation in a trapped ion system might give us a new perspective to understand neutrinos and the electro-weak interaction.
Analog quantum simulators aim at using controllable systems to simulate systems that are hard to access in an experiment. Some examples of the analog quantum simulation are as follows: a black hole is simulated in Bose--Einstein condensation by sonic gravitation correspondence \cite{Garay2000Sonic}; the Klein paradox is simulated using two ions \cite{Gerritsma2011Quantum}; the Majorana equation and unphysical operations are simulated using a trapped ion \cite{Casanova2012Quantum}; and quantum field theory is simulated using trapped ions\cite{Casanova2011Quantum,Xiang2018Experimental}. The Zitterbewegung phenomenon, which is a trembling motion for quantum-relativistic particles, including Weyl fermions, has been widely discussed in the recent years\cite{lamata2007dirac,Gerritsma2010Quantum,Barut1984The,Guertin1973Zitterbewegung,Rusin2010Zitterbewegung,Qu2013Observation}. For example, the theoretical proposal and the experimental result of the quantum simulation of the Dirac equation and Zitterbewegung have been provided by \cite{lamata2007dirac,Gerritsma2010Quantum}. The neutrino's Zitterbewegung has been discussed by \cite{Barut1984The}.

We present herein a method of simulating the $1+2$-dimensional Weyl equation using the analog quantum simulator. The two-dimension Zitterbewegung of Weyl fermions can be tested by the trapped ion experiment. Another interesting phenomenon that can be tested through the experiment is the axisymmetric Weyl fermion initial state that will evolve into a non-axisymmetric state. We show that the Weyl equation evolution can be simulated in a single trapped ion using red- and blue-sideband (bichromatic) laser beams from the $x$- and $y$-directions, respectively. The initial state is prepared by Doppler cooling, sideband cooling, and displacement Hamiltonian. The average momentum of the simulated particle is adjusted by the displacement Hamiltonian. We can use the wave function reconstruction method \cite{Gerritsma2010Quantum} to obtain the average position and possibility distribution at a certain time $t$ by measuring the excited state's population.

\section{Effective Hamiltonian, initial state preparation, and evolution}
\label{sec:1}
The Weyl equation is written as follows:
\begin{equation}
i \hbar \frac{\partial \Psi_{L}}{\partial t}=-c(\sigma_{x}\hat{p}_{x}+\sigma_{y}\hat{p}_{y}+\sigma_{z}\hat{p}_{z})\Psi_{L},
\end{equation}
where $\sigma_{i} (i=x,y,z)$ are Pauli matrices; $c$ is the {speed of light}; and $\Psi_{L}$ is the state of the left-handed two-component Weyl fermions.

\begin{figure}
\centering
\includegraphics[width=80 mm]{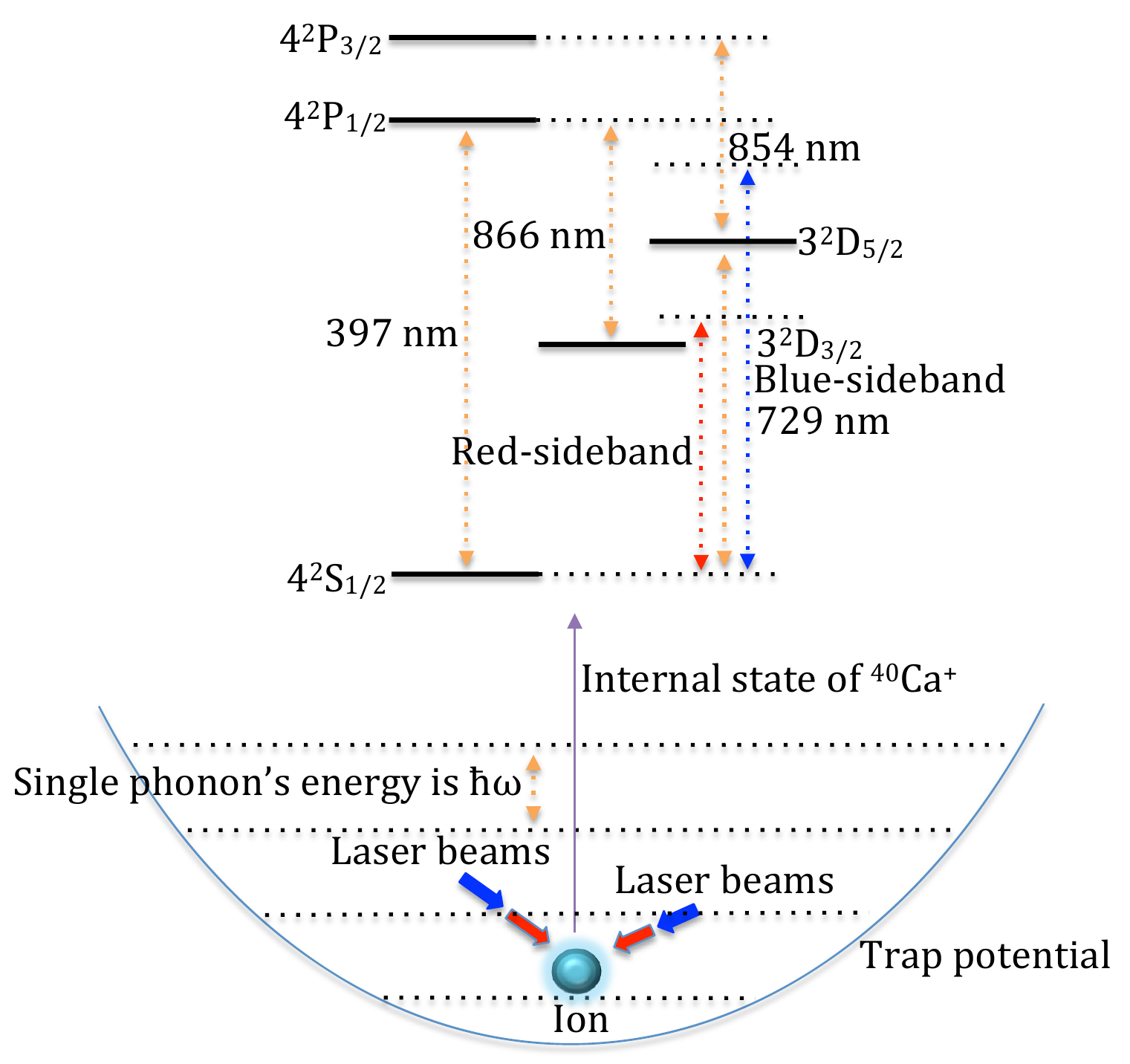} 
\caption{A single trapped ion ($^{40}Ca^{+}$) and its energy level diagrams of external and internal states are shown. The external state of the ion is determined by the number of phonons. We choose two internal states $S_{1/2}, D_{5/2}$ of the ion as the computational basis $|0\rangle, |1\rangle$ of qubit, where $S_{1/2}$ is the ground state of the internal state of the ion, and $D_{5/2}$ is a metastable state (approximately 1 s lifetime) of the ion. After Doppler cooling and sideband cooling, the bichromatic laser beams from the x- and y-directions are used one after another to prepare the initial state. The bichromatic laser beams from the $x$- and $y$-directions are synchronously used to simulate the Weyl equation Hamiltonian \eqref{Weyl}. In the measurement stage, the bichromatic laser beams in the $y$-direction are used to simulate the displacement Hamiltonian.
	}
	\label{fig0}
\end{figure}
We consider an ion trapped in the linear Paul trap to simulate the Weyl equation. For simplicity and to not lose generality, we choose the $^{40}Ca^{+}$-ion as an example. After Doppler cooling and sideband cooling, the trap potential of the ion is treated as a harmonic potential, and the external state of the ion is treated as a quantum harmonic oscillator. The external state of the ion is quantum state and determined by the number of phonons. Each phonon's energy is $\hbar \omega$ (Fig. \ref{fig0}). We choose two internal states of the $^{40}Ca^{+}$-ion as $|0\rangle$, $|1\rangle$ and $|0\rangle=S_{1/2}$, $|1\rangle=D_{5/2}$. The 729 nm, 854 nm and 866 nm laser beams are used to initialize the internal state of $^{40}Ca^{+}$ into $|0\rangle$. The 729 nm laser beam can pump $|0\rangle$ into $|1\rangle$ and realize any possible internal state
\begin{eqnarray}
\alpha|0\rangle +\beta |1\rangle,
\end{eqnarray}
where $|\alpha|^{2}+|\beta|^{2}=1$. The 397 nm laser beam is used in the electron shelving method \cite{leibfried2003quantum} to detect $|\beta|^{2}$. The simulated spinor state is defined as follows:
\begin{equation}
\Psi_{L}=\psi(x,y)(\alpha |0\rangle +\beta|1\rangle)=\left(\begin{array}{c}
\psi(x,y)\alpha\\
\psi(x,y)\beta
\end{array}\right),
\end{equation}
where $\psi(x,y)$ represents the external state of the ion and
\begin{equation}
(|\alpha|^{2}+|\beta|^{2})\int dxdy |\psi(x,y)|^{2}=1.
\end{equation}

One can perform the coupling between the external and internal states of the ion by the red- and blue-sideband laser beams. The red- and blue-sideband photons’ energies are equal to the 729 nm photon's energy minus and plus a phonon's energy, respectively. The Hamiltonian of Jaynes--Cummings (JC) interaction (or {as} we call, red-sideband excitation) is used, which reads as follows \cite{leibfried2003quantum,haffner2008quantum}:
\begin{equation}
H_{r}=\hbar \eta \Omega(a \sigma^{+}e^{i\phi_{r}}+a^{\dagger} \sigma^{-} e^{-i\phi_{r}}),
\end{equation}
where $\eta=k\sqrt{\hbar/2m\omega}$ is the Lamb--Dicke parameter; $m$ is the ion's mass; $a^{\dagger}$ and $a$ are the creation and annihilation operators of phonons (acting on external state), respectively; $\Omega$ is the coupling strength; $k$ is the wave number of the laser field \cite{lamata2007dirac}; $\phi_{r}$ is the phase of the red-sideband laser beam; $\sigma^{+}$ and $\sigma^{-}$ are the operators acting on the internal state $\alpha|0\rangle+\beta|1\rangle$ and $\sigma^{\pm}=\frac{1}{2}(\sigma_{x}\pm i\sigma_{y})$.
The Hamiltonian of the anti-Jaynes--Cummings (AJC) interaction (or as we call, blue-sideband) is also used, which reads as follows:
\begin{equation}
H_{b}=\hbar\eta\Omega(a^{\dagger}\sigma^{+}e^{i\phi_{b}}+a \sigma^{-} e^{-i\phi_{b}}),
\end{equation}
where $\phi_{b}$ is the phase of the blue-sideband laser beam. The momentum and position operators of the ion are presented as follows
\begin{equation}
\hat{p}=-i\sqrt{\frac{m \omega \hbar}{2}} (a-a^{\dagger}), \hat{x}=\sqrt{\frac{ \hbar}{2m \omega}} (a+a^{\dagger}).
\end{equation}

We apply the bichromatic laser fields in the $x$-direction and $y$-direction synchronously. The effective Hamiltonian can be obtained as follows by setting $\phi_{r,x}=\frac{\pi}{2}$, $\phi_{b,x}=-\frac{\pi}{2}$, $\phi_{r,y}=0$, $\phi_{b,y}=\pi$, $\Omega_{x}=\Omega_{y}=\Omega$ \cite{lamata2007dirac,Gerritsma2010Quantum}
\begin{equation}\label{Weyl}
H_{D}=-c(\sigma_{x}\hat{p}_{x}+\sigma_{y}\hat{p}_{y}),
\end{equation}
where $c:=\hbar \eta\Omega\sqrt{\frac{2}{m\omega\hbar}}$ is the speed of light.
Here, only the $\sigma_{z}\hat{p}_{z}$ term is omitted. It is still interesting to know when the simulated equation is the $1+2$-dimension Weyl equation because it involves the two-dimensional Zitterbewegung and an interesting phenomenon of Weyl fermions. These two novel phenomena of the Weyl fermions in the $1+2$-dimension cannot be shown in the $1+1$-dimension Dirac equation simulation\cite{Gerritsma2010Quantum}. Quantum simulation of the 1+3-dimension Weyl equation is interesting because we can study the dynamics of Weyl fermion more close to real neutrinos, for example, the three-dimensional Zitterbewegung and the difference between right- and left-handed neutrinos. Particularly, for 1+3 dimensional Weyl fermions, numerical methods involve considerable computational difficulties.

\begin{figure}
\centering
\includegraphics[width=80 mm]{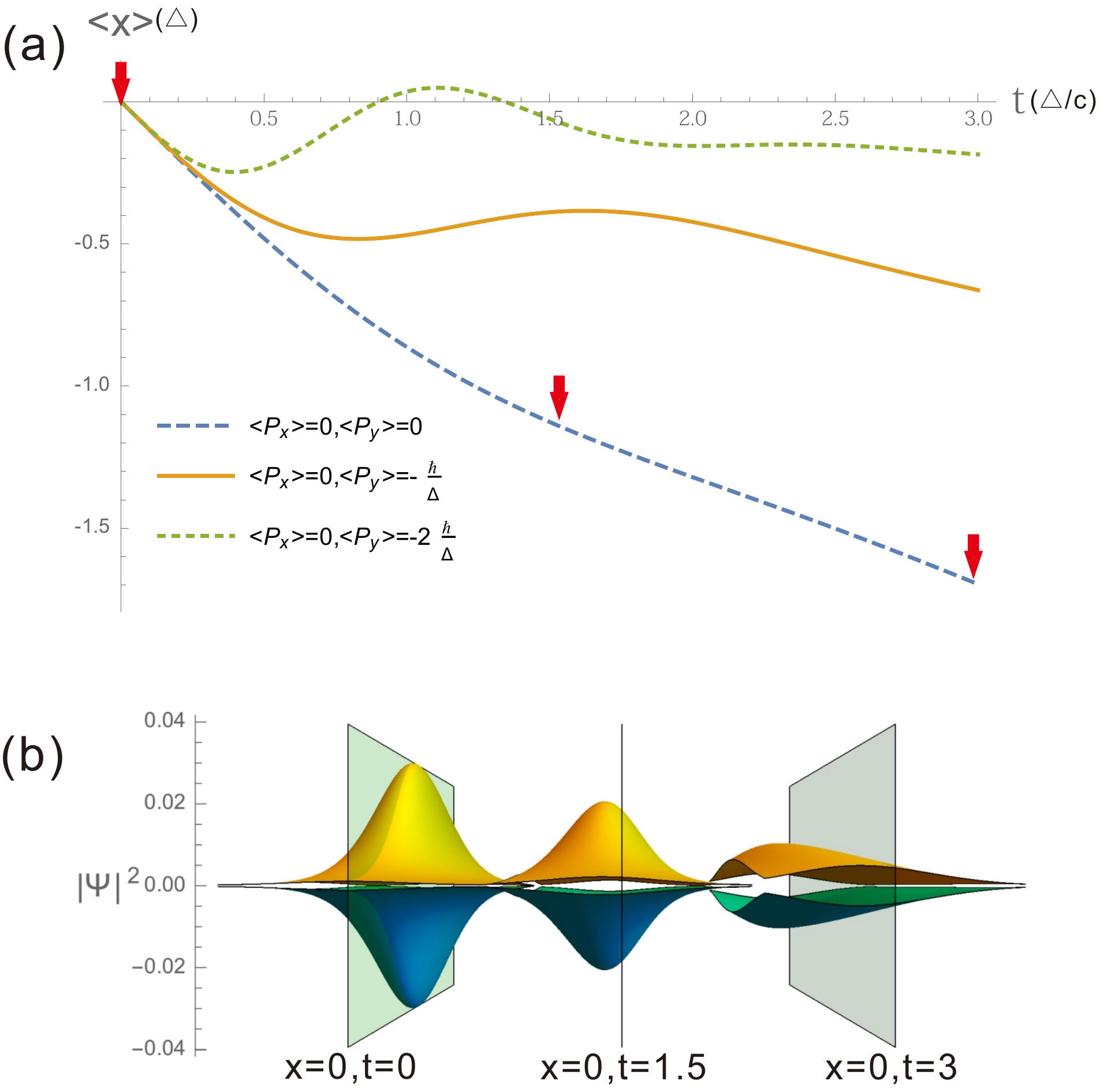} 
	\caption{In graphics (a), the average position, $\langle \hat{x}(t) \rangle$, is shown, where $\langle \hat{p}_{x} \rangle=0$,$\langle \hat{p}_{y} \rangle=0, -\frac{\hbar}{\Delta}, -\frac{2\hbar}{\Delta} $ and $\frac{\Delta}{c}=\frac{1}{2\eta \Omega}$. The corresponding average position $\langle \hat{y}(t) \rangle =0$. In graphics (b), the two-dimensional possibility distributions of $|\Psi_{+}(x,y)|^{2}$ and $|\Psi_{-}(x,y)|^{2}$ are shown ($|\Psi_{-}(x,y)|^{2}$ is inverted), where $-4\Delta \leqslant x,y\leqslant 4\Delta$, $\langle \hat{p}_{x} \rangle=0$, $\langle \hat{p}_{y} \rangle=0$ and $t=0, 1.5, 3$. 
The blue-line trajectory is especially interesting because the initial state corresponding to the blue line is axisymmetric in position space and momentum space, but the evolved state is not axisymmetric. The corresponding $\langle \hat{y}(t) \rangle =0$, but $\langle \hat{x}(t) \rangle \neq 0$.}
	\label{fig1}
\end{figure}

\begin{figure}
\centering
\includegraphics[width=80 mm]{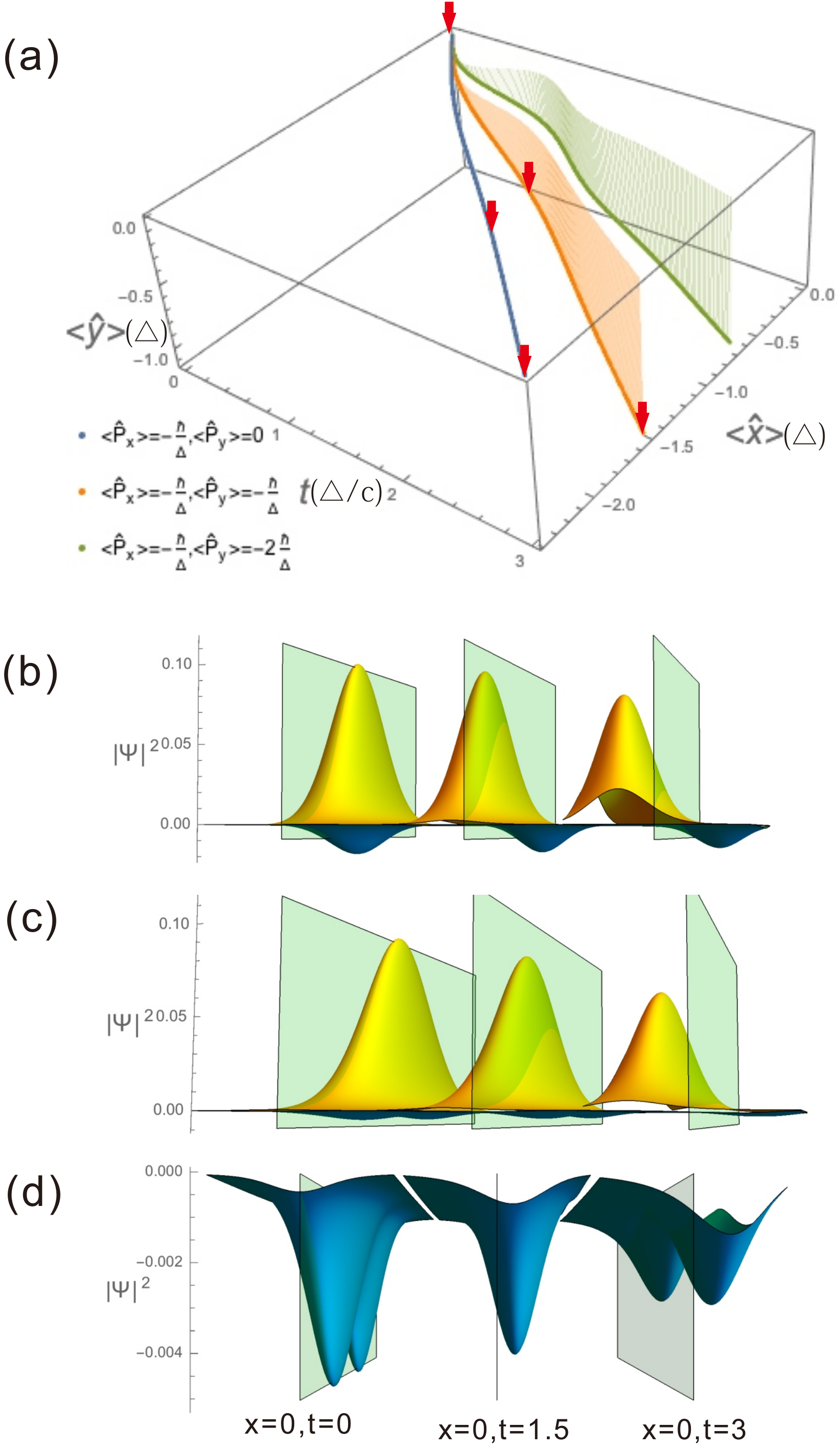} 
\caption{In graphics (a), the average position of $\langle \vec{\hat{r}}(t) \rangle$ is shown, where $\langle \hat{p}_{x}\rangle=-\frac{\hbar}{\Delta}$,$\langle \hat{p}_{y} \rangle=0, -\frac{\hbar}{\Delta}, -\frac{2\hbar}{\Delta}$. In graphics (b), the two-dimensional possibility distributions of $|\Psi_{+}(x,y)|^{2}$ and $|\Psi_{-}(x,y)|^{2}$ are shown, where $-4\Delta \leqslant x,y\leqslant 4\Delta$, $\langle \hat{p}_{x} \rangle=-\frac{\hbar}{\Delta}$, $\langle \hat{p}_{y} \rangle=-\frac{\hbar}{\Delta}$ and $t=0, 1.5, 3$. The positive and negative energy wave functions are separate; hence, the Zitterbewegung fades away. In graphics (c), almost pure positive energy wave packets with the average momentum $\langle \hat{p}_{x} \rangle=-\frac{\hbar}{\Delta}$,$\langle \hat{p}_{y} \rangle=0$ at $t=0, 1.5, 3$, where the inverted $|\Psi_{-}|^{2}$ is too small to be seen, are shown. The corresponding blue line in graphics (a) is an almost straight line, and the speed is close to the speed of light. In graphics (d), the inverted negative energy possibility distributions $|\Psi_{-}|^{2}$, which cannot be seen in graphics (c) have two peaks.}
	\label{fig2}
\end{figure}

One has to prepare the initial state before simulating the evolution of the Weyl equation. First, Doppler cooling and sideband cooling are needed, and the external state in the $x$- and $y$-directions should be cooled to ground state (phonon's number equals 0). The external state is a Gauss wave function, while the internal state can be chosen such that
\begin{eqnarray}\label{initialstate}
\Psi_{L}(t=0,x,y)=\frac{1}{2\sqrt{\pi}\Delta} e^{-\frac{x^{2}+y^{2}}{4\Delta^{2}}}\left(\begin{array}{c} 1\\
1\end{array}\right).
\end{eqnarray}
Here, $\int_{0}^{\infty}e^{-x^{2}}dx=\frac{\sqrt{\pi}}{2}$ is used, and $\Delta=\sqrt{\frac{\hbar}{2m \omega}}$. Second, the external and internal states can be further adjusted. We add the bichromatic laser beams and set $\phi_{r}=\phi_{b}=0$. The effective Hamiltonian is achieved as follows:
\begin{eqnarray}
\tilde{H}_{D}=\frac{\hbar \eta \Omega}{\Delta} \hat{x}\sigma_{x},
\end{eqnarray}
where $\tilde{H}_{D}$ is the displacement Hamiltonian. If the bichromatic laser beams act for a period $t=\frac{n}{\eta\Omega}$, the Gauss wave function \eqref{initialstate} will be altered to
\begin{eqnarray}\label{wavefunction}
\frac{1}{2\sqrt{\pi}\Delta} e^{\frac{-i }{\Delta}nx}e^{-\frac{x^{2}+y^{2}}{4\Delta^{2}}}\left(\begin{array}{c} 1\\
1\end{array}\right).
\end{eqnarray}
We use the position representation and eigenstate of $\sigma_{x}$ with eigenvalue $1$ such that $\hat{x}\rightarrow x$ and $\sigma_{x}\rightarrow 1$. Similarly, the bichromatic laser beams can be added to the $y$-direction and the wave function can be achieved {as follows}:
\begin{eqnarray}\label{wavefunctionxy}
\frac{1}{2\sqrt{\pi}\Delta} e^{\frac{-i }{\Delta}nx}e^{\frac{-i }{\Delta}my}e^{-\frac{x^{2}+y^{2}}{4\Delta^{2}}}\left(\begin{array}{c} 1\\
1\end{array}\right),
\end{eqnarray}
where $m,n\geqslant0$. The initial wave functions in Figs. \ref{fig1} and \ref{fig2} are described by Eq.~\eqref{wavefunctionxy} with a particular $m,n$. The average momenta of \eqref{wavefunctionxy} are
\begin{eqnarray}
\langle \hat{p}_{x}\rangle=-\frac{n\hbar}{\Delta}, \ \ \langle \hat{p}_{y}\rangle=-\frac{m\hbar}{\Delta}.
\end{eqnarray}

The wave packet solution of the Weyl equation is the superposition state of the positive and negative energy solutions with different momenta.
\begin{eqnarray}\label{Psi}
\Psi_{L}&=&\Psi_{+}+\Psi_{-} \\ \nonumber
&=&\int \left(\begin{array}{c}
-\frac{(p_{x}-i p_{y})}{\sqrt{p_{x}^{2}+p_{y}^{2}}}\\ \nonumber
1
\end{array}
\right)\chi_{+}(p_{x},p_{y}) e^{\frac{i}{\hbar}(p_{x}x+p_{y}y-Et)}dp_{x}dp_{y}\\
&&+\int \left(\begin{array}{c}
\frac{(p_{x}-i p_{y})}{\sqrt{p_{x}^{2}+p_{y}^{2}}}\\
1
\end{array}
\right)\chi_{-}(p_{x},p_{y}) e^{\frac{i}{\hbar}(p_{x}x+p_{y}y+Et)}dp_{x}dp_{y}, \nonumber
\end{eqnarray}
where $E=c\sqrt{p_{x}^{2}+p_{y}^{2}}$.

If the initial state is described as \eqref{wavefunctionxy}, $\chi_{+}$ and $\chi_{-}$ are
\begin{subequations}
\begin{eqnarray}\label{chione} \nonumber
\chi_{+}(p_{x},p_{y})&=&\frac{\Delta}{4\hbar^{2}\pi\sqrt{\pi}}\left(1-\frac{p_{x}+ip_{y}}{\sqrt{p_{x}^{2}+p_{y}^{2}}}\right)e^{-\frac{\Delta^{2}}{\hbar^{2}}(p_{x}+\frac{n\hbar}{\Delta})^{2}}\\
&&e^{-\frac{\Delta^{2}}{\hbar^{2}}(p_{y}+\frac{m\hbar}{\Delta})^{2}},\\ \label{chitwo} \nonumber
\chi_{-}(p_{x},p_{y})&=&\frac{\Delta}{4\hbar^{2}\pi\sqrt{\pi}}\left(1+\frac{p_{x}+ip_{y}}{\sqrt{p_{x}^{2}+p_{y}^{2}}}\right)e^{-\frac{\Delta^{2}}{\hbar^{2}}(p_{x}+\frac{n\hbar}{\Delta})^{2}}\\
&&e^{-\frac{\Delta^{2}}{\hbar^{2}}(p_{y}+\frac{m\hbar}{\Delta})^{2}}.
\end{eqnarray}
\end{subequations}
Substituting Eqs.~\eqref{chione} and \eqref{chitwo} into Eq.~\eqref{Psi}, the wave function at $t$, $\Psi_{L}(t,x,y)$ can be derived. Figs. \ref{fig1}b and \ref{fig2} (graphics (b) and (c)) show the possibility distributions, $|\Psi_{+}|^{2}$ and $|\Psi_{-}|^{2}$ (inverted), in different average momenta at $t=0, 1.5, 3$. In Fig. \ref{fig1}b, the overlap between $\Psi_{+}$ and $\Psi_{-}$ always exists, leading to the Zitterbewegung. The wave packets in Fig. \ref{fig1} move in the $x$-direction, but not in the $y$-direction even though the initial wave packet is axisymmetric. Fig. \ref{fig2} (graphics (b) and (c)) shows that the overlap reduces or even disappears over time, leading to fading of the Zitterbewegung. The wave packet in Fig. \ref{fig2}b has movement in both {$x$-} and $y$-directions. Fig. \ref{fig2}c shows almost pure positive energy wave packets. The negative energy wave packets are too small to be seen. Fig. \ref{fig2}d shows the inverted $|\Psi_{-}|^{2}$ with $\langle \hat{p}_{x} \rangle=-\frac{\hbar}{\Delta}$, $\langle \hat{p}_{y} \rangle=0$. Two peaks are illustrated in Fig. \ref{fig2}d. 

In the Heisenberg picture, the position operator of the Weyl fermions as a function of time is presented as follows:
\begin{equation}\label{Zitter}
\vec{\hat{r}}(t)=\vec{\hat{r}}(0)+c^{2}\vec{\hat{p}}H^{-1}_{D}t+\frac{i}{2}\hbar c H^{-1}_{D}(e^{2i H_{D}t/\hbar}-1)(\vec{\sigma}(0)+c\vec{\hat{p}}H^{-1}_{D}),
\end{equation}
where $\vec{\sigma}(0)$ is the $\vec{\sigma}$ operator's initial state and $c^{2}\vec{\hat{p}}H^{-1}_{D}$ is a classical velocity operator. The eigenvalues of the classical velocity operator are $\pm \frac{c \vec{p}}{\sqrt{p_{x}^{2}+p_{y}^{2}}}$. The magnitude of the classical velocity is the speed of light. The last term in Eq.~\eqref{Zitter} is the Zitterbewegung term. Without loss of generality, $\vec{\hat{r}}(0)=(i\hbar \frac{d}{dp_{x}},i\hbar \frac{d}{dp_{y}})$  and we set $\vec{\sigma}(0)=(\sigma_{x},\sigma_{y})$.

The momentum representation of the wave function \eqref{wavefunctionxy} is
\begin{eqnarray}\label{psiL}
\psi_{L}=\frac{\Delta}{\sqrt{\pi}\hbar}e^{-\frac{{\Delta}^{2}}{\hbar^{2}}(p_{x}+\frac{n\hbar}{\Delta})^{2}}e^{-\frac{{\Delta}^{2}}{\hbar^{2}}(p_{y}+\frac{m\hbar}{\Delta})^{2}}\left(\begin{array}{c}  1\\1   \end{array}\right).
\end{eqnarray}
The average value of the operator $\vec{\hat{r}}(t)$ is
\begin{eqnarray}\label{average}
\langle \vec{\hat{r}}(t)\rangle=\int_{-\infty}^{\infty}\int_{-\infty}^{\infty}dp_{x}dp_{y}\psi_{L}^{\dagger}\vec{\hat{r}}(t)\psi_{L}.
\end{eqnarray}
The explicit $\langle \vec{\hat{r}}(t)\rangle$ is
\begin{eqnarray}  \label{averagex}
\langle \hat{x}(t)\rangle&=&\frac{\Delta^{2}}{\pi \hbar^{2} }\int_{-\infty}^{\infty}\int_{-\infty}^{\infty}dp_{x}dp_{y}\frac{-1}{p_{x}^{2}+p_{y}^{2}}
\left(2c p_{x}^{2}t \right.\\  \nonumber
&+&\left.\frac{\hbar p_{y}^{2} sin (\frac{2c \sqrt{p_{x}^{2}+p_{y}^{2}}t}{\hbar})}{\sqrt{p_{x}^{2}+p_{y}^{2}}}\right)
e^{-\frac{2\Delta^{2}}{\hbar^{2}}(p_{x}+\frac{n\hbar}{\Delta})^{2}}e^{-\frac{2\Delta^{2}}{\hbar^{2}}(p_{y}+\frac{m\hbar}{\Delta})^{2}},\\  
\label{averagey}
\langle \hat{y}(t)\rangle&=&\frac{\Delta^{2}}{\pi \hbar^{2} }\int_{-\infty}^{\infty}\int_{-\infty}^{\infty}dp_{x}dp_{y}\frac{-1}{p_{x}^{2}+p_{y}^{2}}\left(2cp_{x}p_{y}t \right. \\  \nonumber
&-&\left. \frac{\hbar p_{x}p_{y} sin (\frac{2c \sqrt{p_{x}^{2}+p_{y}^{2}}t}{\hbar})}{\sqrt{p_{x}^{2}+p_{y}^{2}}}\right)
e^{-\frac{2\Delta^{2}}{\hbar^{2}}(p_{x}+\frac{n\hbar}{\Delta})^{2}}e^{-\frac{2\Delta^{2}}{\hbar^{2}}(p_{y}+\frac{m\hbar}{\Delta})^{2}}.
\end{eqnarray}

Figures \ref{fig1}a and \ref{fig2}a show the average position $\langle\vec{\hat{r}}(t) \rangle$ of the Weyl fermions with different average momenta, which represent $2$-dimensional Zitterbewegung behavior. In Fig. \ref{fig1}a, the destructive interference between different momentum components leads to $\langle \hat{y}(t)\rangle=0$, while the constructive interference leads to $\langle \hat{x}(t)\rangle\neq 0$. Eq.~\eqref{average} shows that, when $n=0$ or $m=0$, $\langle \hat{y}(t)\rangle$ is the integral of an odd function, so the corresponding $\langle \hat{y}(t)\rangle=0$. When $n=0,m=1$, Eqs.~\eqref{averagex}, \eqref{averagey} give us the phenomenon of the wavefunction's displacement in the $y$-direction, but Zitterbewegung occurs in the $x$-direction. In fact, this is a case of coincidence and the Zitterbewegung term always exists in $\langle \hat{x}\rangle$ and $\langle \hat{y}\rangle$ if $|\Psi_{+}|^{2}$ and $|\Psi_{-}|^{2}$ overlap. Even though the magnitude of the velocity is equal to the speed of light for each positive and negative energy solution, the average velocity of the wave packet is always lower than the speed of light because the wave packet includes parts with opposite direction velocity. When $m=1,n=0$, the corresponding wavefunction is composed of almost purely positive energy wave packets such that its velocity magnitude approaches the speed of light and its trajectory approaches a straight line.

Figures \ref{fig1} and \ref{fig2} show that, when $|\Psi_{+}|^{2}$ and $|\Psi_{-}|^{2}$ overlap, the Zitterbewegung phenomenon exists. Zitterbewegung also disappears when the overlap disappears. Therefore, Zitterbewegung originates from the interference of positive and negative energy wave functions. Furthermore, the higher the ratio 
\begin{eqnarray}
\frac{\int dx dy |\Psi_{+}|^{2}}{ \int dx dy |\Psi_{-}|^{2}}>1,
\end{eqnarray}
the closer the average velocity of the whole wave packet is to the speed of light. Figures \ref{fig1} and \ref{fig2} also show that the average position of the simulated Weyl fermion not only depends on the interference of positive and negative energy wave functions, but also on the interference of different momentum components. The interference between different momentum components evolves the axisymmetric wave function into a non-axisymmetric one. This interesting phenomenon of a single Weyl fermion might enhance our understanding of neutrinos.

\section{Measurement method}
\label{sec6}
We will provide herein a method \cite{Gerritsma2010Quantum,wallentowitz1995reconstruction} to test the Zitterbewegung and possibility distribution in the experiment. We apply the bichromatic laser beams in the $y$-direction only and set $\phi_{r}=\phi_{b}=0$. The effective Hamiltonian is then presented as
\begin{equation}
H_{disp}=(H_{r}+H_{b})_{y}=\frac{\hbar \eta\Omega}{\Delta}\hat{y} \sigma_{x}.
\end{equation}
The unitary operator is
\begin{equation}
U=exp(-\frac{i}{\hbar}H_{disp}t)=e^{-ik\hat{y}\sigma_{x}/2},
\end{equation}
where $k=\frac{2\eta\Omega}{\Delta} t$.

The observable quantity is
\begin{equation}
O(k)=U^{\dagger}\sigma_{z}U=cos(k\hat{y})\sigma_{z}+sin(k\hat{y})\sigma_{y}.
\end{equation}
We define $|+\rangle_{z}$ as the eigenstate of $\sigma_{z}$ with eigenvalue $+1$ and $|+\rangle_{y}$ as the eigenstate of $\sigma_{y}$ with eigenvalue $+1$. We obtain the following if the initial state is the eigenstate of $\sigma_{z}$:
\begin{eqnarray} \nonumber
&&\langle \psi(x,y)|\langle +|_{z}O(k)|+\rangle_{z}|\psi(x,y)\rangle\\  
&=&\langle cos(k\hat{y})\rangle=2\langle \psi(y)|{|\rho_{z}(t,\hat{y})|}^{2}|\psi(y)\rangle-1.
\end{eqnarray}
Similarly, we have
\begin{eqnarray} \nonumber
&&\langle \psi(x,y)|\langle +|_{y}O(k)|+\rangle_{y}|\psi(x,y)\rangle\\
&=&\langle sin(k\hat{y})\rangle=2\langle \psi(y)|{|\rho_{y}(t,\hat{y})|}^{2}|\psi(y)\rangle-1,
\end{eqnarray}
where $\rho_{z}(t,\hat{y})=\langle e|+\rangle_{z}$, $\rho_{y}(t,\hat{y})=\langle e|+\rangle_{y}$ and $\langle e|= \langle 1|U$. $\langle \psi(y)|{|\rho_{z}(t,\hat{y})|}^{2}|\psi(y)\rangle$ and $\langle \psi(y)|{|\rho_{y}(t,\hat{y})|}^{2}|\psi(y)\rangle$ are the quantities that can be directly measured by the electron shelving method. $\langle \psi(y)|{|\rho_{z}(t,\hat{y})|}^{2}|\psi(y)\rangle$ ($\langle \psi(y)|{|\rho_{y}(t,\hat{y})|}^{2}|\psi(y)\rangle$) is the average population possibility of the excitation state $|1\rangle$ with the initial state $|+\rangle_{z}$ ($|+\rangle_{y}$). 
The exponential function of $k\hat{y}$ is obtained as follows:
\begin{equation}
\langle e^{ik\hat{y}}\rangle=\langle cos(k\hat{y})+i sin(k\hat{y})\rangle.
\end{equation}
The Fourier transformation of $\langle e^{ik\hat{y}}\rangle$ is presented as
\begin{equation}
\int _{0}^{\infty} \langle e^{ik\hat{y}}\rangle e^{-i2\pi t y_{0}}dt \approx\langle\psi(y_{0})|\delta({y}^{\prime}-y_{0})|\psi(y_{0})\rangle=|\psi({y}^{\prime})|^{2},
\end{equation}
where $y^{\prime}=\frac{\eta\Omega}{\pi\Delta}y$ (in position representation, $\hat{y}\rightarrow y$). Similarly, $|\psi({x}^{\prime})|^{2}$ can be obtained. The wave function reconstruction method is shown. The average position of the simulated particle can be calculated as follows:
\begin{eqnarray}
\langle \hat{x} \rangle=\int |\psi(x)|^{2}x dx, \langle \hat{y} \rangle=\int |\psi(y)|^{2}y dy.
\end{eqnarray}
\section{Conclusion}
\label{con}
We herein proposed a method for simulating a $1+2$-dimension Weyl equation using a single trapped ion. The phenomenon of the two-dimensional Zitterbewegung of the Weyl fermions is discussed. Zitterbewegung originates from the interference of $\Psi_{+}$ and $\Psi_{-}$. Even though the initial state of the Weyl fermions is axisymmetric in the two-dimensional position space and momentum space, the interference between the different momentum wave functions leads to zero motion in the $y$-direction and nonzero motion in the $x$-direction and evolves the axisymmetric wave function into a non-axisymmetric one. 
Studying the dynamics of a single particle to multiple particles, transitions from relativistic to non-relativistic states, and changing from spin-$0$ to spin-$\frac{1}{2}$ and spin-$1$ states, either theoretically or experimentally is interesting.

\begin{acknowledgements}
This work was supported by the National Basic Research Program of China under Grant No. 2016YFA0301903 and the National Natural Science Foundation of China under Grant Nos. 11174370, 11304387, 61632021, 11305262, 61205108, and 11574398. 
\end{acknowledgements}



\bibliographystyle{apsrev4-1}

\renewcommand{\baselinestretch}{1}

\end{document}